\documentclass[twocolumn]{aastex6}
\usepackage{natbib}
\usepackage{amsmath}

\begin{document}
\title{Methanol formation via oxygen insertion chemistry in ices}
\author{Jennifer B. Bergner\altaffilmark{1}, Karin I. \"Oberg\altaffilmark{2}, Mahesh Rajappan\altaffilmark{2}}
\altaffiltext{1}{Harvard University Department of Chemistry and Chemical Biology, 10 Oxford Street, Cambridge, MA 02138, USA}
\altaffiltext{2}{Harvard-Smithsonian Center for Astrophysics, 60 Garden Street, Cambridge, MA 02138, USA}

\begin{abstract}
We present experimental constraints on the insertion of oxygen atoms into methane to form methanol in astrophysical ice analogs.  In gas-phase and theoretical studies this process has previously been demonstrated to have a very low or non-existent energy barrier, but the energetics and mechanisms have not yet been characterized in the solid state.  We use a deuterium UV lamp filtered by a sapphire window to selectively dissociate O$_2$ within a mixture of O$_2$:CH$_4$ and observe efficient production of CH$_3$OH via O($^1$D) insertion.  CH$_3$OH growth curves are fit with a kinetic model, and we observe no temperature dependence of the reaction rate constant at temperatures below the oxygen desorption temperature of 25K.  Through an analysis of side products we determine the branching ratio of ice-phase oxygen insertion into CH$_4$: $\sim$65\% of insertions lead to CH$_3$OH with the remainder leading instead to H$_2$CO formation.  There is no evidence for CH$_3$ or OH radical formation, indicating that the fragmentation is not an important channel and that insertions typically lead to increased chemical complexity.  CH$_3$OH formation from O$_2$ and CH$_4$ diluted in a CO-dominated ice similarly shows no temperature dependence, consistent with expectations that insertion proceeds with a small or non-existent barrier.  Oxygen insertion chemistry in ices should therefore be efficient under low-temperature ISM-like conditions, and could provide an important channel to complex organic molecule formation on grain surfaces in cold interstellar regions such as cloud cores and protoplanetary disk midplanes.
\end{abstract}


\section{Introduction}
\label{sec_intro}
Complex organic molecules (COMs) have been detected towards star-forming regions at all stages of evolution, including molecular clouds, protostellar hot cores, envelopes, and outflows, and protoplanetary disks \citep[e.g.][]{Blake1987,Bottinelli2004,Arce2008,Oberg2010,Oberg2015}.  It is of great interest to understand the rich chemistry that feeds the formation and destruction of these molecules in the interstellar medium (ISM) in order to constrain the chemical inventories available for pre-biotic chemistry as solar systems develop.  To explain COM production, current astrochemical models typically rely on grain-surface radical recombination chemistry that becomes active in lukewarm ($\sim$30K) ices \citep{Garrod2008,Herbst2009}.  However, observations of COMs towards very cold interstellar environments such as pre-stellar cores \citep[][e.g.]{Oberg2010,Bacmann2012,Cernicharo2012} indicate that a cold pathway to complex molecule formation must also be active.  

A central challenge to building chemical complexity at low temperatures is the incorporation of several heavy elements into molecules.  To address this, many experimental studies have focused on the hydrogenation of unsaturated oxygen-bearing molecules.  Repeated hydrogenation of CO has been shown to be efficient and leads to the production of the stable molecules H$_2$CO and CH$_3$OH \citep{Watanabe2002,Fuchs2009}:
\begin{equation}
\mathrm{CO \xrightarrow[(1)]{H} HCO \xrightarrow[(2)]{H} H_2CO \xrightarrow[(3)]{H} H_3CO \xrightarrow[(4)]{H} CH_3OH}
\label{COhydrog}
\end{equation}
In addition, H atom bombardment of CO ices has been recently shown to form even more complex molecules: H atom abstractions along the CO hydrogenation pathway (\ref{COhydrog}) lead to enhanced populations of radical species, which can recombine to form larger COMs \citep{Fedoseev2015a,Chuang2015}.  However, this recombination chemistry requires diffusion of relatively heavy radical species, which in interstellar ices are not expected to be mobile at low temperatures. 

Here, we present experiments demonstrating an oxygen insertion mechanism as an alternative pathway to COM formation at low temperatures, using the test case of CH$_3$OH formation from O insertion into CH$_4$.  Unlike radical recombination pathways, oxygen insertion does not require diffusion of heavy species.  The ability to convert hydrocarbons directly to COMs represents a novel channel to explain observed COM abundances.

The gas-phase insertion of excited O($^1$D) into CH$_4$ has been well-studied: initially, vibrationally excited CH$_3$OH is formed, and at sufficiently high pressures can be collisionally stabilized; otherwise, the molecule fragments into the radical products CH$_3$ + OH \citep{DeMore1967,Lin1973}.  In the gas phase, therefore, insertions are typically net destructive and lead to smaller rather than larger molecules.  Gas-phase oxygen insertion has been shown experimentally to be essentially barrierless \citep[e.g.][]{DeMore1967}, and indeed theoretical studies suggest a small $\sim$280K barrier \citep{Yu2004}.  

Oxygen insertion has also been qualitatively demonstrated in condensed systems.  In \citet{Appelman1989}, HOF was photolyzed within a CH$_4$ matrix under high vacuum (10$^{-7}$ Torr) conditions.  \citet{Parnis1993} and \citet{Lugez1994} both studied oxygen insertion into CH$_4$ within Ar matrices, using photolysis of N$_2$O and O$_3$ respectively for O($^1$D) atom generation.   Each of these studies qualitatively demonstrated CH$_3$OH formation via oxygen insertion in the solid state.  However, to date there is no quantitative or mechanistic description of this process.   Moreover, the ice compositions in these previous studies are not astrophysically realistic.  In order to evaluate the extent to which oxygen insertion can lead to chemical complexity in the ISM, a detailed understanding of how this process occurs in astrophysical ice analogs is required.  In particular, it is essential to determine the energetic feasibility under ISM-like conditions, and the efficiency of CH$_3$OH formation compared to fragmentation and other product formation. 

O($^1$D) atoms, which are required for this insertion process, are in the first electronically excited state with energies $\sim$2eV higher than the ground state O($^3$P).  O($^1$D) production has been demonstrated by UV photolysis of a number of molecules which should be abundant in interstellar ice mantles, including O$_2$, CO$_2$, O$_3$, and H$_2$O, as well as electron impact of O$_2$ and CO$_2$ \citep{Lee1977,Slanger1971,DeMore1966, Stief1975,Cosby1993,Kedzierski2013}.  CO$_2$ and H$_2$O are of particular importance, as H$_2$O is the main constituent of interstellar ices and CO$_2$ is typically present at abundances of $\sim$10-30\% with respect to H$_2$O \citep{Oberg2011}.  Photolysis of CO$_2$ follows the dissociation channel to O($^1$D) + CO($^1\Sigma^+$) between 120nm and 170nm, with measured efficiencies of 94\% and 100\% at 157nm and 147nm respectively \citep{Zhu1989,Slanger1971}.  From 105nm-145nm, H$_2$O dissociates to H$_2$ + O($^1$D) with an efficiency of $\sim$10\% \citep{Ung1974,Stief1975,Slanger1982}.  Importantly, for both CO$_2$ and H$_2$O, O($^1$D) can be generated upon exposure to Lyman-$\alpha$ irradiation (121.6nm), which dominates the UV spectrum in cloud cores and protoplanetary disks \citep[e.g.][]{Oberg2016}.  Therefore, dissociation of common oxygen-bearing molecules in ice mantles should produce these excited oxygen atoms, which may then insert into neighboring hydrocarbons within the ice.  

To assess the energetics and mechanism of this reaction scheme in astrophysical ice analogs, we selectively dissociate O$_2$ within mixed O$_2$:CH$_4$ ices and O$_2$:CH$_4$:CO ices.  
In Section \ref{sec_exp}, we describe the experimental apparatus and procedures.  Section \ref{sec_analysis} details the data analysis techniques used, and Section \ref{sec_results} presents the results.  In Section \ref{sec_discussion} we discuss the reaction network and mechanisms in this system, as well as the astrophysical implications of this process.

\section{Experimental Details}
\label{sec_exp}
We use the ulta-high vacuum experiment described in detail in \citet{Lauck2015}.  The chamber is evacuated to a base pressure of $\sim$ 5 x 10$^{-10}$ Torr.  A closed-cycle He cryostat cools a CsI substrate window to temperatures as low as 9K.  Temperature is monitored by a temperature controller (LakeShore 335) with an estimated accuracy of 2K and a relative uncertainty of 0.1K.  Ices are typically grown on the substrate by introduction of gases through a 4.8mm diameter dosing pipe at 0.7 inches from the substrate.  For co-deposition experiments, two separate dosing pipes at $\sim$1.2 inches from the substrate were used to introduce gases.  Experiments were performed using the following gases: CH$_4$ (99.9\% purity, Aldrich), $^{13}$CH$_4$ (99\%, Aldrich), CD$_4$ (99\%, Aldrich), $^{16}$O$_2$ (99.98\%, Aldrich), $^{18}$O$_2$ (97\%, Aldrich), Ar (99.95\%, Aldrich), and CO (99.95\%, Aldrich).  To obtain mixtures, gases were combined in a differentially pumped gas line with a base pressure $\sim$10$^{-4}$ Torr.  A Fourier transform infrared spectrometer (Bruker Vertex 70v) in transmission mode was used to measure infrared-active molecules in the ice.  A quadrupole mass spectrometer (Pfeiffer QMG 220M1) continuously monitored the gas-phase species present in the chamber.  

\begin{figure}
	\centering
	\includegraphics[width=0.9\linewidth]{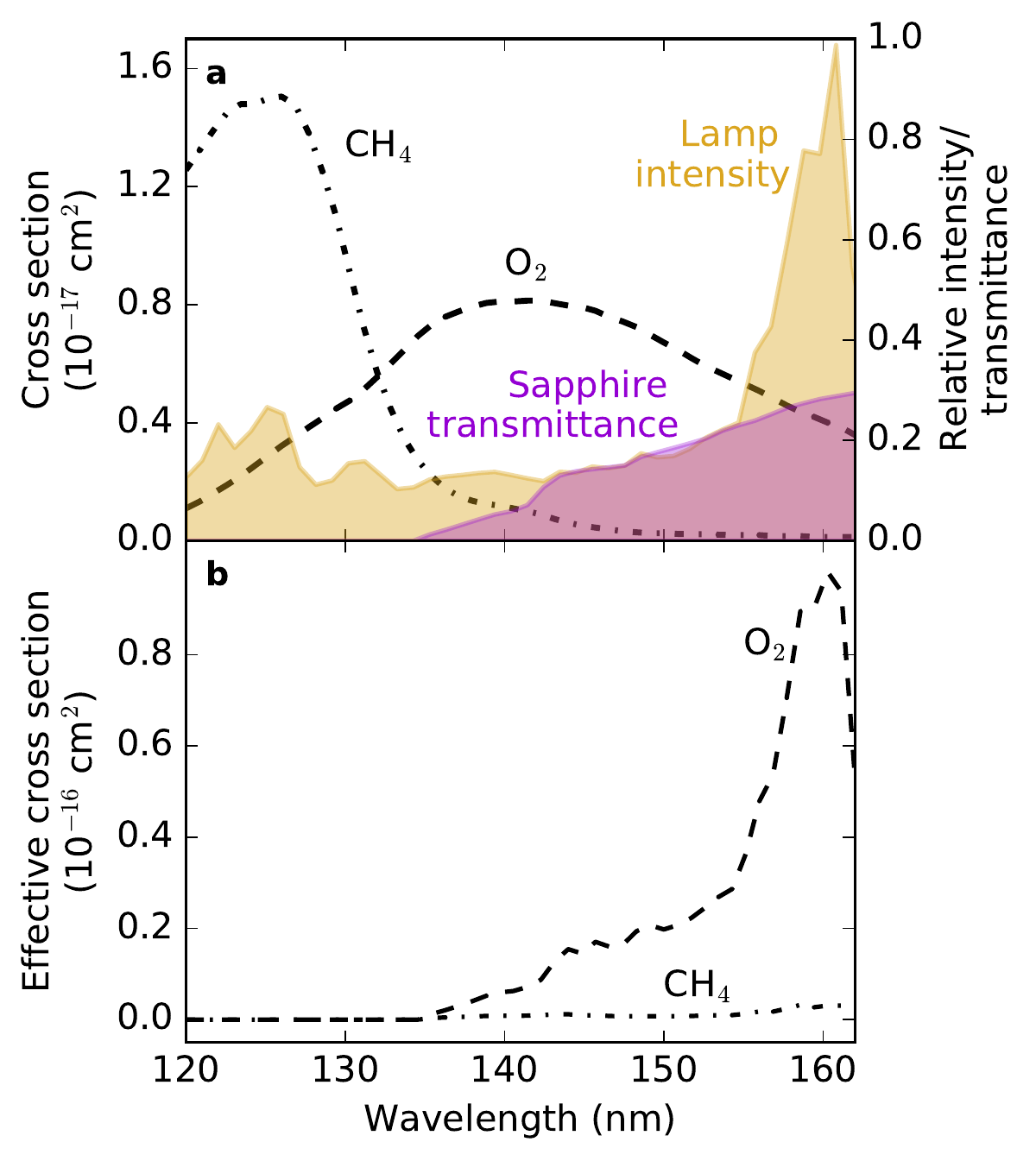}
	\caption{a: CH$_4$ (black dash-dot) and O$_2$ (black dash) UV absorption cross-sections \citep{Cruz-Diaz2014a} overlaid with the H$_2$D$_2$ UV lamp output spectrum (gold) and sapphire transmission spectrum (purple).  b: The effective absorption cross-sections for CH$_4$ and O$_2$ resulting from the convolution of each absorption cross-section with the relative lamp intensity and window transmittance spectra.}
	\label{crsec}
\end{figure}

To selectively dissociate O$_2$ within a CH$_4$:O$_2$ mixture, we irradiate the ice samples with a H$_2$D$_2$ lamp (Hamamatsu L11798) filtered by a 0.08" thick sapphire window (MPF Products, Inc.)  Figure \ref{crsec}a shows the spectral distribution of the UV lamp overlaid with the transmittance spectrum of sapphire with a similar thickness\footnote{http://resources.montanainstruments.com/help/window-materials} and the UV absorption cross-sections for both CH$_4$ and O$_2$ \citep{Cruz-Diaz2014a}.  The convolution of lamp intensity, sapphire transmittance, and absorption cross-section for each species is shown as the ``effective'' cross-section in Figure \ref{crsec}b.  This represents the relative absorption of each molecule under the irradiation conditions of the present experiment.  The absorption of CH$_4$ is negligible due to the cutoff of the sapphire window, while O$_2$ still has a large absorption profile which is dominated by the 160.8nm Lyman band of the H$_2$D$_2$ lamp.  The photoproducts of O$_2$ in this wavelength regime are well-characterized: upon exposure to light at wavelengths from 140-175nm, O$_2$ dissociates into O($^3$P) + O($^1$D) with an efficiency of unity \citep{Lee1977}.

All experimental details are summarized in Table 1.  We define different groups of experiments, each varying with respect to the Fiducial category as indicated: (I- Fiducial) $^{16}$O$_2$:$^{13}$CH$_4$ $\sim$1.4:1, 50ML total ice thickness; (II) 20ML thickness; (III) $\sim$0.2:1 ratio; (IV) capping layer of argon; (V) 90ML thickness; (VI): $^{12}$CD$_4$:$^{18}$O$_2$; (VII) other isotopologues (qualitative analysis only); (VIII) CO:$^{13}$CH$_4$:$^{16}$O$_2$ 4:1:1, 125ML thickness; (IX): CO:$^{13}$CH$_4$:$^{16}$O$_2$ 7:1:1, 170ML thickness; (X): control experiments- only 1 reactant.

The experimental procedure for groups I-VII involved a single deposition of a mixture of CH$_4$ and O$_2$; for groups VIII and IX, separate dosing tubes were used to co-deposit CO and a CH$_4$:O$_2$ mixture.  In all cases, dosing was performed at 9K, followed by a 2 hour sample irradiation at a set temperature at an angle of 45$^{o}$ by the H$_2$D$_2$ lamp.  During irradiation, IR scans were taken every 3 minutes.  Following irradiation, a temperature programmed desorption (TPD) was performed by ramping the sample temperature at a rate of 2K/min to 200K.  During the temperature ramp, IR scans were taken every 2 minutes and desorbing species were monitored with the QMS.

\begin{deluxetable*}{lccrrc} 
	\tabletypesize{\footnotesize}
	\tablecaption{Experiment summary}
	\tablecolumns{6} 
	\tablewidth{\linewidth} 
	\tablehead{\colhead{Experiment}                            	&
		\colhead{Group}										&
		\colhead{Irradiation}                 			  	&
		\colhead{Ice composition}                             		         			 	&
		\colhead{\hspace{0.2in} Ratio\hspace{0.2in} }							&
		\colhead{Total thickness }                           \\
		\colhead{} 												&
		\colhead{}							  		&
		\colhead{temp (K)}						  	&
		\colhead{}							  			&
		\colhead{}							  		&
		\colhead{(ML)}                             								  	}		
	\startdata
1 & I & 9 & $^{16}$O$_2$ : $^{13}$CH$_4$ & 1.3 : 1 & 53\\ 
2 & I & 9 & $^{16}$O$_2$ : $^{13}$CH$_4$ & 1.8 : 1 & 52\\ 
3 & I &14 & $^{16}$O$_2$ : $^{13}$CH$_4$ & 1.4 : 1 & 53\\ 
4 & I &19 & $^{16}$O$_2$ : $^{13}$CH$_4$ & 1.3 : 1 & 51\\ 
5 & I &24 & $^{16}$O$_2$ : $^{13}$CH$_4$ & 1.4 : 1 & 50\\ 
6 & I &25 & $^{16}$O$_2$ : $^{13}$CH$_4$ & 1.4 : 1 & 54\\ 
7 & II & 9 & $^{16}$O$_2$ : $^{13}$CH$_4$ & 0.9 : 1 & 18\\ 
8 & II &14 & $^{16}$O$_2$ : $^{13}$CH$_4$ & 1.6 : 1 & 23\\ 
9 & II &19 & $^{16}$O$_2$ : $^{13}$CH$_4$ & 1.3 : 1 & 19\\ 
10 & III & 9 & $^{16}$O$_2$ : $^{13}$CH$_4$ & 0.2 : 1 & 26\\ 
11 & III &19 & $^{16}$O$_2$ : $^{13}$CH$_4$ & 0.2 : 1 & 23\\ 
12 & III &24 & $^{16}$O$_2$ : $^{13}$CH$_4$ & 0.2 : 1 & 23\\ 
13\tablenotemark{a} & IV & 9 & $^{16}$O$_2$ : $^{13}$CH$_4$ & 1.2 : 1 & 38\\ 
14 & V & 9 & $^{16}$O$_2$ : $^{13}$CH$_4$ & 1.6 : 1 & 92\\ 
15 & V & 9 & $^{16}$O$_2$ : $^{13}$CH$_4$ & 1.6 : 1 & 87\\ 
16\tablenotemark{a} & VI & 9 & $^{18}$O$_2$ : $^{12}$CD$_4$ & 1.5 : 1 & 37\\ 
17 & VI & 9 & $^{18}$O$_2$ : $^{12}$CD$_4$ & 1.9 : 1 & 66\\ 
18 & VI &19 & $^{18}$O$_2$ : $^{12}$CD$_4$ & 1.7 : 1 & 59\\ 
19\tablenotemark{a} & VII & 9 & $^{18}$O$_2$ : $^{13}$CH$_4$ & 0.8 : 1 & 32\\ 
20 & VII & 9 & $^{16}$O$_2$ : $^{12}$CH$_4$ & 1.5 : 1 & 55\\ 
21 & VIII & 9 & $^{12}$CO : $^{16}$O$_2$ : $^{13}$CH$_4$ & 4.3 : 1.1 : 1 & 125\\ 
22 & VIII &19 & $^{12}$CO : $^{16}$O$_2$ : $^{13}$CH$_4$ & 4.5 : 1.1 : 1 & 127\\ 
23 & VIII &19 & $^{12}$CO : $^{16}$O$_2$ : $^{13}$CH$_4$ & 4.4 : 1.1 : 1 & 125\\ 
24 & IX & 9 & $^{12}$CO : $^{16}$O$_2$ : $^{13}$CH$_4$ & 6.9 : 1.0 : 1 & 172\\ 
25 & IX &14 & $^{12}$CO : $^{16}$O$_2$ : $^{13}$CH$_4$ & 6.8 : 1.0 : 1 & 172\\ 
26 & IX &19 & $^{12}$CO : $^{16}$O$_2$ : $^{13}$CH$_4$ & 7.0 : 1.0 : 1 & 174\\ 
27 & X & 9 & $^{13}$CH$_4$ & - & 15\\ 
28 & X & 9 & $^{16}$O$_2$ \hspace{0.42in} & - & 23\\ 
\enddata
\tablenotetext{a}{with Ar top layer}
\tablenotetext{}{See text for Group designations}
\label{expts}
\end{deluxetable*}

\section{Data analysis}
\label{sec_analysis}
\subsection{O$_2$ thickness determination}
Since O$_2$ is infrared inactive, we quantify the O$_2$ dose by calibrating the time-integrated QMS signal to the IR-determined CH$_4$ dose as in \citet{Fayolle2013}:
\begin{equation}
N_{O_2} = N_{CH_4}\frac{\int I_{O_2^+}dt}{\int I_{CH_4^+}dt}\frac{\sigma_{CH_4^+}}{\sigma_{O_2^+}}.
\end{equation}
Here, $N_x$ is the column density of species $x$, $\int I_xdt$ is the time-integrated QMS intensity of each molecule's dominant ion, and $\sigma_x$ is the gas-phase electron impact ionization cross-section of each molecular ion at 70eV.  The QMS mass signals $m/z$ 17, 20, 32, and 36 are used to trace $^{13}$CH$_4$, CD$_4$, $^{16}$O$_2$, and $^{18}$O$_2$ respectively; these masses should not be contaminated by signal from any other molecule.  O$_2$ and CH$_4$ cross-sections are taken from \citet{Straub1996} and \citet{Straub1997} respectively. The average value of $N_{CH_4}/\int I_{CH_4^+}dt$ from all experiments was used to convert to O$_2$ doses; based on variations between experiments, we expect an O$_2$ dose uncertainty of about 30\%.  The CH$_4$ thickness uncertainty is $\sim$20\%, arising mainly due to the uncertainty in band strength.

\subsection{UV flux}
\label{UV_flux}
The photon flux from the UV lamp was measured with a NIST calibrated AXUV-100G photo-diode at the sample holder to be $\sim$1.4 x 10$^{13}$ photons cm$^{-2}$ s$^{-1}$, with a measurement uncertainty of $\sim$5\% for the wavelengths of interest in this work.  For a 2-hour irradiation, this results in a total fluence of 1 x 10$^{17}$ cm$^{-2}$.  To ensure that UV photons penetrate the entire ice sample, we calculate the attenuation of photons for experiments of different ice thickness.  The photon attenuation is calculated at 160.8nm, since this represents the peak of the ``effective'' cross-section (Section \ref{sec_exp}), using the formula
\begin{equation}
N_{X} = -\frac{1}{\sigma_{X}(\lambda)}ln\frac{\rm{I_t}(\lambda)}{\rm{I_0}(\lambda)}
\end{equation} 
where $N_{X}$ is the column density of molecule X (molecule cm$^{-2}$), $\sigma_{X}(\lambda)$ is the UV absorption cross-section, and $\rm{I_t}(\lambda)$ and $\rm{I_0}(\lambda)$ are the transmitted and incident intensities, respectively.  We assume the standard monolayer coverage of 10$^{15}$ molecules cm$^{-2}$ and cross-sections $\sigma_{O_2}(160.8\mathrm{nm})$ = 3.9x10$^{-18}$ cm$^{2}$ and $\sigma_{CO}(160.8\mathrm{nm})$ = 0.9x10$^{-18}$ cm$^2$ \citep{Cruz-Diaz2014, Cruz-Diaz2014a}.  For O$_2$ ice, 15, 35, and 65ML will absorb 6\%, 13\%, and 22\% of photons respectively.  For CO ice, 85 and 135ML will absorb 7\% and 11\% of photons respectively.  Therefore, photon attenuation is small and should not impact the experimental results even for the thickest O$_2$ and CO ices.

\subsection{IR spectra and growth curves}
\label{quant}
IR spectra were used to determine the initial ice column densities of CH$_4$ as well as the growth of products during irradiation.  Each spectrum is averaged over 128 interferograms and takes approximately 2 minutes to complete.  Column densities of each species were calculated using the integrated area of IR features:
\begin{equation}
N_{i} = \frac{2.3\int\tau_{i}(\nu)d\nu}{A_{i}} \label{col_density},
\end{equation}
where $N_{i}$ is column density (molecule cm$^{-2}$), $\int\tau_{i}(\nu)d\nu$ is the integrated area of the IR band (absorbance units), and $A_{i}$ is the band strength in optical depth units.  The standard monolayer coverage of 10$^{15}$ molecules cm$^{-2}$ was assumed.

\begin{deluxetable*}{llccl} 
	\tabletypesize{\footnotesize}
	\tablecaption{IR band strengths}
	\tablecolumns{5} 
	\tablewidth{0.95\linewidth} 
	\tablehead{\colhead{Molecule}                        	&
		\colhead{Mode}							         	&
		\multicolumn{2}{c}{Line center (cm$^{-1}$)}      				&
		\colhead{$A$}					                   	\\
		\colhead{}											&
		\colhead{}											&
		\colhead{($^{13}$C/$^{16}$O/H)}					&
		\colhead{($^{12}$C/$^{18}$O/D)}					&
		\colhead{(cm molec$^{-1}$)}							}
	\startdata
CH$_4$	&	$\nu_4$ bend        &   1291  	& 989	&	8.4 x 10$^{-18}$ ($a$)	\\
CH$_3$OH	&	$\nu_8$ CO str	&	1008  	& 952	&	1.8 x 10$^{-17}$ ($a$)	\\
O$_3$	&	$\nu_3$ a-str    	&	1034  	& 975	&   8.8 x 10$^{-18}$ ($b$)	\\
CO      &   str			    	&	2092	&		&	1.1 x 10$^{-17}$ ($a$)	\\
CO$_2$  &   $\nu_3$ a-str		&	2276	&		&	1.3 x 10$^{-16}$ ($a$)	\\
H$_2$CO &   $\nu_2$ CO str		&	1683	&		&	9.6 x 10$^{-18}$ ($a$)	\\
        &   $\nu_3$ CH$_2$ sc. 	&   1498   	&		&	3.9 x 10$^{-18}$ ($a$) \\
H$_2$O  &   $\nu_2$ bend    	&   1652    &		& 	1.1 x 10$^{-17}$ ($a$)  \\
\enddata
	\tablenotetext{a}{\citet{Bouilloud2015}}
	\tablenotetext{b}{\citet{Loeffler2006}; re-scaled from RAIRS value}
	\label{bands}
\end{deluxetable*}

Product growth curves can only be measured for $^{13}$CH$_4$:$^{16}$O$_2$ and $^{12}$CD$_4$:$^{18}$O$_2$ mixtures (Groups I-VI and VIII-IX), as other isotopologue combinations have overlapping O$_3$ and CH$_3$OH infrared features.  IR band centers and strengths used for quantifying molecule abundances are listed in Table 2.  

O$_3$ column densities are highly uncertain since only RAIRS band strengths were available in the literature.  However, absolute O$_3$ measurements are not required for quantifying the reaction between O atoms and CH$_4$.  To estimate the ozone abundance, we assume that the ratio between RAIRS and transmission band strengths is consistent within an experiment \citep[e.g. ][]{Ioppolo2008}.  We therefore scale the RAIRS band strength of O$_3$ reported in \citet{Loeffler2006} by the ratio of the transmission and RAIRS band strengths for the H$_2$O $\nu_2$ bending mode \citep[$\sim$1.4;][]{Bouilloud2015,Loeffler2006}.

For CH$_3$OH, the direct product of O insertion into CH$_4$, the strongest feature is the $\nu_8$ band.  Due to the proximity of the O$_3$ $\nu_3$ band, both features were fit simultaneously with Gaussian profiles.  Each feature is well-fit by a double Guassian; thus, for each spectrum we fit four Gaussians as seen in Figure \ref{fits}a.  Fitting is done using the python $emcee$ package \citep{Foreman-Mackey2013}.  In addition to the four Gaussian profiles, we simultaneously fit a local linear baseline term as part of the overall fit.  This enables us to incorporate any uncertainties that arise from the choice of baseline into subsequent analysis.  Figure \ref{corner} in Appendix A shows an example corner plot displaying the degree of parameter covariance as well as the fit convergence. 

\subsection{Kinetic modeling}
\label{kinetics}
The reaction steps leading to methanol formation are:
\begin{eqnarray}
\mathrm{O}_2 \xrightarrow[h\nu]{k_{pd}} \mathrm{O}(^3\mathrm{P}) + \mathrm{O}(^1\mathrm{D})
\label{rxn_odiss} \\
\mathrm{O}(^1\mathrm{D}) + \mathrm{CH}_4 \xrightarrow{k_r} \mathrm{CH_3OH}
\label{rxn_ins}
\end{eqnarray}
where k$_{pd}$ is the photodissociation rate of O$_2$ and $k_r$ is the rate of oxygen insertion to form methanol.  In addition to formation via O insertion, CH$_3$OH can also be photo-dissociated by UV light in this wavelength range.  To account for potential CH$_3$OH loss we include a destruction term in the kinetic model:
\begin{equation}
\mathrm{CH_3OH} \xrightarrow[h\nu]{k_{dest}} \mathrm{products}.
\label{rxn_mdiss}
\end{equation}
As detailed in previous work, there are numerous possible products of CH$_3$OH processing \citep{Bennett2007a,Oberg2009a}. For the purposes of fitting growth curves, we are mainly interested in the loss of CH$_3$OH, and discuss the potential formation of other products in more detail in Section \ref{sec_results}.  We note that CH$_3$OH could also be consumed by other mechanisms such as H abstractions, however we assume that photodissociation is the dominant destruction mechanism.  Incorporating other pathways would require modeling the entire system simultaneously, which is not practical given difficulties in measuring column densities of each product throughout the experiment.

From Reactions \ref{rxn_ins} and \ref{rxn_mdiss} we obtain the integrated rate law describing CH$_3$OH formation:
\begin{equation}
\mathrm{[CH_3OH]}(t) = \frac{k_r\mathrm{N_{ss}}(e^{-k_{dest}t} - e^{-k_rt})}{k_r - k_{dest}} 
\label{rateloss}
\end{equation}
where N$_{\mathrm{ss}}$ is a proportionality factor representing the steady-state abundance. We have assumed that the timescales of O($^1$D) atom formation and destruction are much faster than the reaction timescale, and therefore the photodissociation kinetics of O$_2$ (Reaction \ref{rxn_odiss}) do not contribute to the overall growth rate.  Indeed, typical CH$_3$OH formation rates are on the order of 10$^{-4}$ML s$^{-1}$, while the timescale for O formation ($\sim$ $\sigma_{O_2}$ x $flux$ x N$_{O_2}$) is an order of magnitude faster, and the rate constant for O($^1$D) relaxation to O($^3$P) in solids is on the order of $\sim$1 s$^{-1}$ \citep{Mohammed1990}.

\begin{figure}
	\centering
	\includegraphics[width=0.9\linewidth]{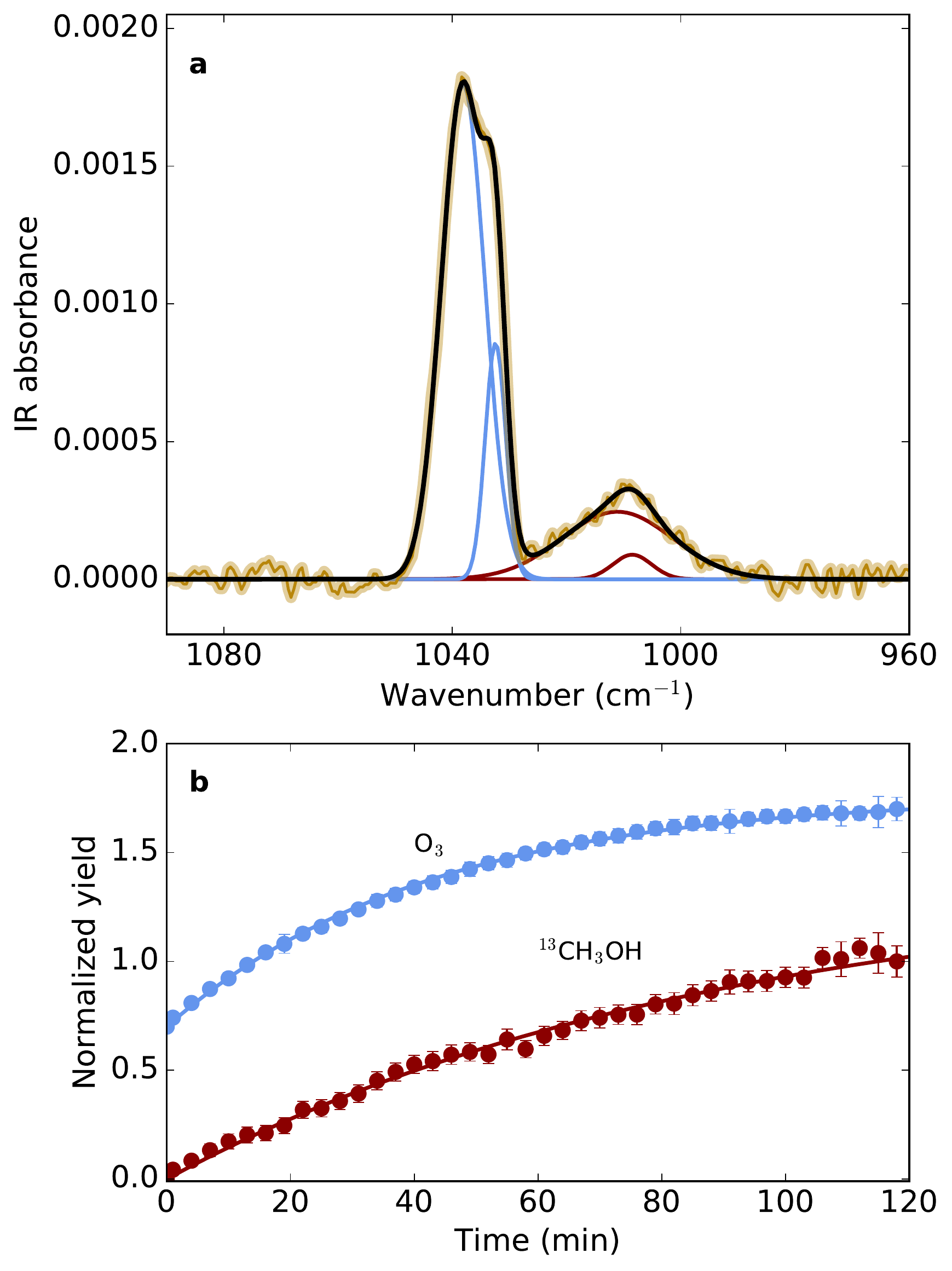}
	\caption{a: Example fit to infrared bands of O$_3$ at 1034cm$^{-1}$ and $^{13}$CH$_3$OH at 1008cm$^{-1}$.  The experimental spectrum is shown in gold and the total fit in black, with individual Gaussians in blue for ozone and red for methanol.  b: Scatter points show experimental growth curves for O$_3$ (blue; offset) and $^{13}$CH$_3$OH (red).  Kinetic fits are shown as solid lines.  Both panels show results from Experiment 3.  Product yield has been normalized by the column density at the end of 2 hours of irradiation.}
	\label{fits}
\end{figure}

For fitting Equation \ref{rateloss} to growth curves, we assume that the rate constant for photodestruction of CH$_3$OH is equal to that of photon absorption, or $\sigma_{\mathrm{CH_3OH}}$ x $flux$.  $\sigma_{CH_3OH}$ is taken from \citet{Cruz-Diaz2014} to be 2.9 x 10$^{-18}$ cm$^2$ at 160.8nm for a total absorption rate = 4.1 x 10$^{-5}$ s$^{-1}$ = $k_{dest}$.  An example fit is shown in Figure \ref{fits}b for $^{13}$CH$_3$OH and O$_3$ growth curves during the irradiation of a $^{13}$CH$_4$:$^{16}$O$_2$ ice mixture (Experiment 3).  The fits are very good, indicating that this model is indeed appropriate and that no additional parameters are required to describe the kinetics.  Growth curves along with best-fit kinetic models for all other experiments are shown in Appendix B.

\section{Results}
\label{sec_results}
\subsection{Proof of concept: CH$_3$OH production}
\begin{figure*}
	\centering
	\includegraphics[width=0.6\linewidth]{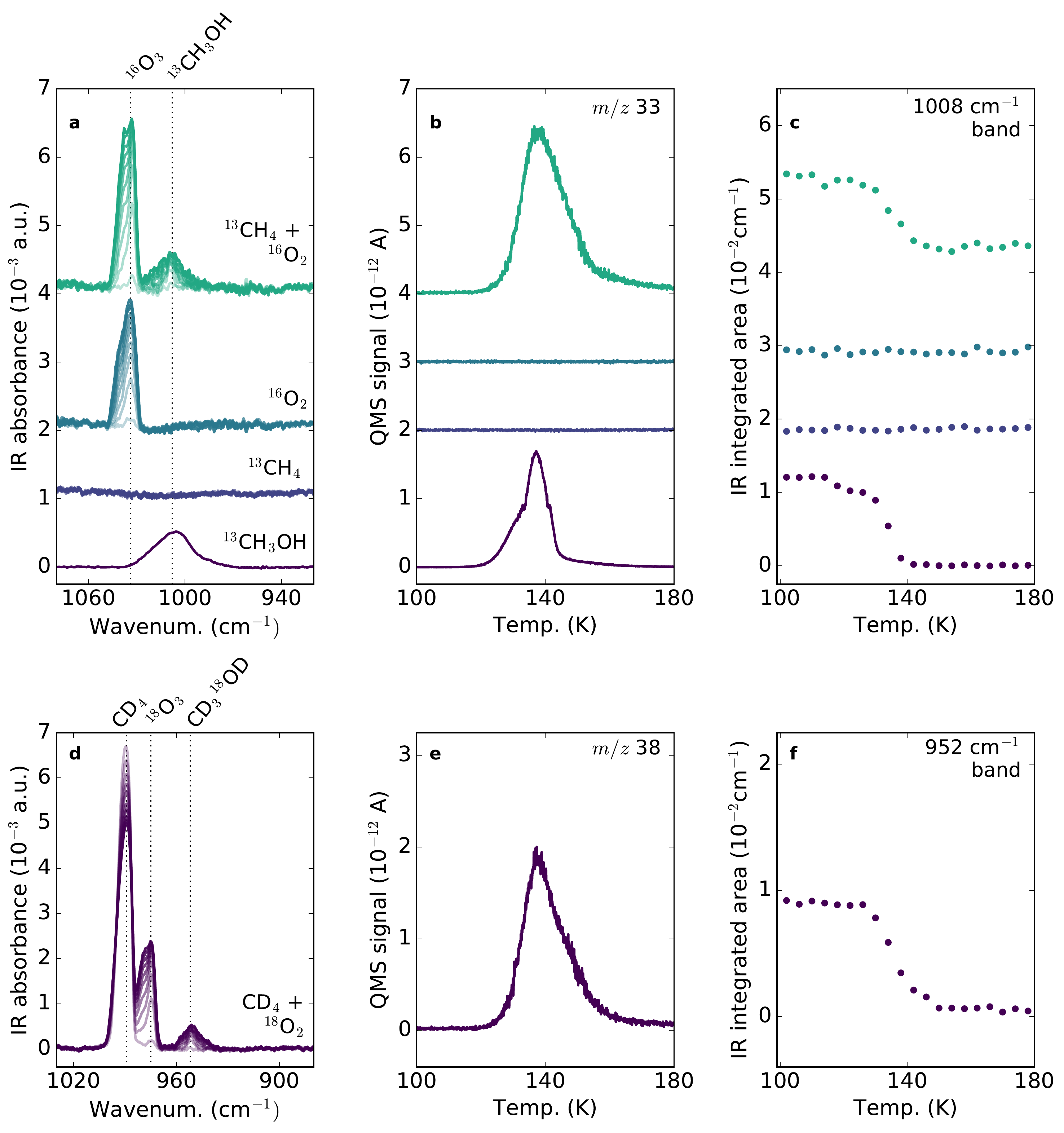}
	\caption{CH$_3$OH production demonstrated through growth of IR features during irradiation (a,d), QMS traces during TPD (b,e), and integrated IR bands during TPD (c,f).  Top panels show results for irradiated $^{13}$CH$_4$ + $^{16}$O$_2$ mixture, irradiated $^{16}$O$_2$ only, irradiated $^{13}$CH$_4$ only, and a $^{13}$CH$_3$OH standard. Bottom panels show results for irradiated CD$_4$:$^{18}$O$_2$.}
	\label{POC}
\end{figure*}

During irradiation of $^{13}$CH$_4$:$^{16}$O$_2$ mixtures, $^{13}$CH$_3$OH formation is observed in situ by the growth of the infrared band at 1008 cm$^{-1}$.  Figure \ref{POC}a shows this feature in the spectrum of pure $^{13}$CH$_3$OH.  The band is not formed during irradiation of pure $^{13}$CH$_4$ or pure $^{16}$O$_2$, but does grow during irradiation of a mixture of $^{13}$CH$_4$:$^{16}$O$_2$.  The growth of $^{16}$O$_3$ can also be seen in the pure $^{16}$O$_2$ and mixed $^{13}$CH$_4$:$^{16}$O$_2$ experiments. 

Figure \ref{POC}b and c show results from the TPD following sample irradiation.  In Figure \ref{POC}b, the QMS trace for $m/z$ 33, which corresponds to the mass of $^{13}$CH$_3$OH, is observed for the mixed irradiated sample.  Desorption occurs around 140K, as in the $^{13}$CH$_3$OH standard.  No $m/z$ 33 signal is observed for either of the pure control experiments.  Figure \ref{POC}c shows the integrated IR band area around 1008cm$^{-1}$ during the temperature ramp.  Again, in the mixed irradiated experiment, the band disappears around 140K and matches well with the $^{13}$CH$_3$OH standard.

Mixtures of CD$_4$:$^{18}$O$_2$ similarly demonstrated growth of CD$_3^{18}$OD.  A standard was not commercially available, however we identify the band at 952cm$^{-1}$ as CD$_3^{18}$OD based on its similar growth to $^{13}$CH$_3$OH during irradiation (Figure \ref{POC}d).  Furthermore, $m/z$ 38 corresponding to CD$_3^{18}$OD desorbs around 140K coincident with the depletion of the IR band at 952cm$^{-1}$(Figure \ref{POC}e and f).  We therefore confirm the O insertion behavior in two isotopologue systems; based on these different lines of evidence, CH$_3$OH is produced during the irradiation of CH$_4$:O$_2$ mixtures.

\begin{deluxetable}{lll} 
	\tabletypesize{\footnotesize}
	\tablecaption{CH$_3$OH growth curve parameter fits: CH$_4$:O$_2$ experiments}
	\tablecolumns{3} 
	\tablewidth{0.75\linewidth} 
	\tablehead{\colhead{Expt.}                           	&
		\colhead{$\mathrm{N_{ss}}$ (ML)}							         	&
		\colhead{$k_r$ (s$^{-1}$)}					        }
	\startdata
1 & 3.5 [0.4] & 1.1 [0.1] x 10$^{-4}$\\ 
2 & 3.0 [0.5] & 1.0 [0.2] x 10$^{-4}$\\ 
3 & 4.9 [1.0] & 6.3 [1.3] x 10$^{-5}$\\ 
4 & 3.1 [0.3] & 1.2 [0.2] x 10$^{-4}$\\ 
5 & 3.8 [0.7] & 8.0 [1.5] x 10$^{-5}$\\ 
6 & 7.1 [1.8] & 4.0 [1.1] x 10$^{-5}$\\ 
7 & 1.5 [0.4] & 1.2 [0.3] x 10$^{-4}$\\ 
8 & 0.8 [0.4] & 1.5 [0.4] x 10$^{-4}$\\ 
9 & 1.1 [0.5] & 1.1 [0.4] x 10$^{-4}$\\ 
10 & 2.5 [0.6] & 7.5 [1.7] x 10$^{-5}$\\ 
11 & 1.2 [0.2] & 1.4 [0.3] x 10$^{-4}$\\ 
12 & 1.0 [0.3] & 1.6 [0.3] x 10$^{-4}$\\ 
13 & 2.1 [0.2] & 1.5 [0.2] x 10$^{-4}$\\ 
14 & 3.8 [1.3] & 9.5 [3.4] x 10$^{-5}$\\ 
15 & 2.8 [0.9] & 1.2 [0.4] x 10$^{-4}$\\ 
16 & 3.1 [1.5] & 7.1 [3.3] x 10$^{-5}$\\ 
17 & 7.0 [3.3] & 3.9 [2.1] x 10$^{-5}$\\ 
18 & 5.7 [2.5] & 5.4 [2.4] x 10$^{-5}$\\ 
	\enddata
	\tablenotetext{}{Uncertainties listed in brackets}
	\label{rates}
\end{deluxetable}

\subsection{O$_2$:CH$_4$ Experiments} 
\subsubsection{Rate constants}
To assess the energetics of oxygen insertion into CH$_4$, experiments were performed at 9K, 14K, 19K, 24K, and 25K.  As described in Section \ref{quant}, first-order kinetic models with photodissociative loss were used to fit each growth curve.  Fits are typically performed for 2 hours of irradiation, however for ices thicker than 30ML fits are performed for 1 hour irradiations since peak overlap between the CH$_3$OH and O$_3$ feature becomes more severe for the higher product column densities.  The parameters $k_r$ and N$_{\mathrm{ss}}$ for each model are listed in Table 3.  The scatter in rate constants from all $^{13}$CH$_4$:$^{16}$O$_2$ experiments performed at 9K was used to estimate uncertainties due to day-to-day systematics.  We find a standard deviation of $\sim$20\% in the rate constant values; for all experiments, this systematics uncertainty was added in quadrature with the growth curve fit errors in order to obtain the final uncertainties on $k_r$.

\begin{figure}
	\centering
	\includegraphics[width=0.8\linewidth]{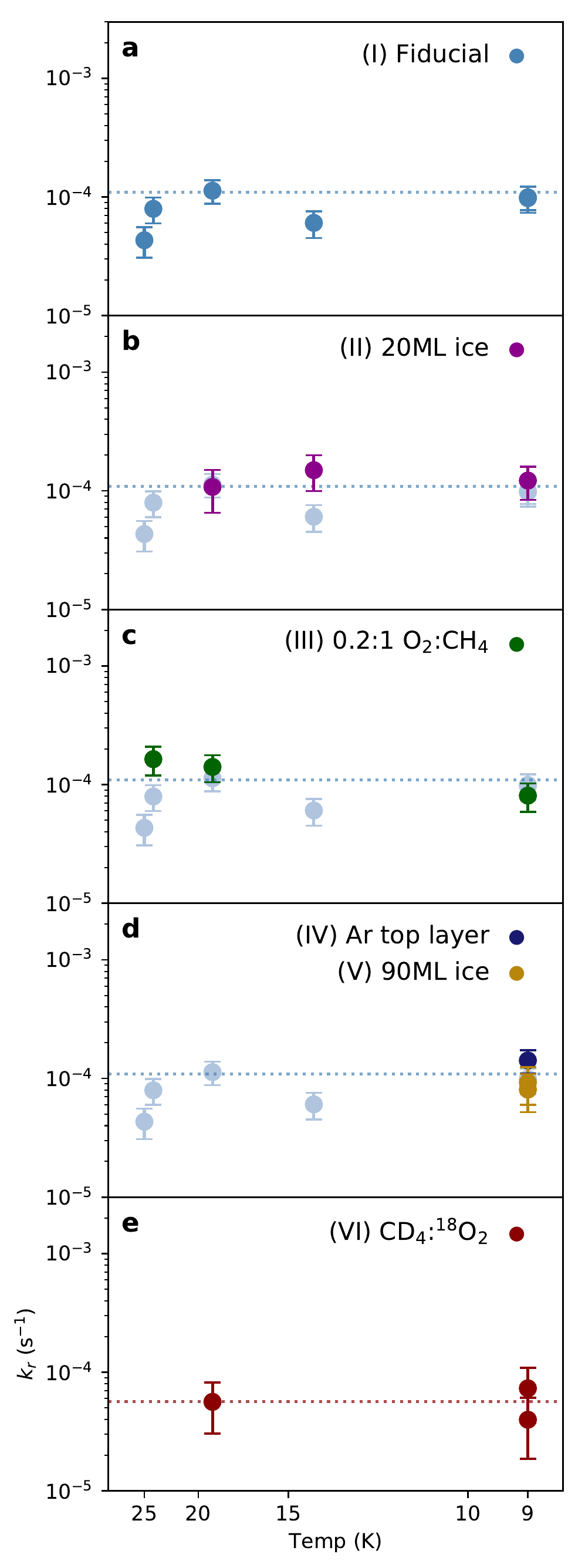}
	\caption{Arrhenius plots for CH$_3$OH growth.  Fiducial experiments ($\sim$1.4:1 $^{13}$CH$_4$:$^{16}$O$_2$, 50ML total ice thickness) are shown in the top panel and reproduced in the other $^{13}$CH$_4$:$^{16}$O$_2$ experiments.  All other panels show experiments that differ from the fiducial group as indicated, with experimental group number shown in parentheses.  Dotted blue lines (a-d) show the mean rate constant for all $^{13}$CH$_4$:$^{16}$O$_2$ experiments; the dotted red line (e) shows the mean rate constant for CD$_4$:$^{18}$O$_2$ experiments .}
	\label{Arrh}
\end{figure}

Figure \ref{Arrh} shows Arrhenius plots of the rate constants for different groups of experiments.  For all $^{13}$CH$_4$:$^{16}$O$_2$ experiments at temperatures below 25K, there is no temperature dependence to the CH$_3$OH formation rate constant: regardless of the ice thickness, the CH$_4$:O$_2$ ratio, or the presence of an inert capping gas, the value of $k_r$ is consistent around 10$^{-4}$s$^{-1}$.  The lower rate at 25K is likely due to the onset of oxygen desorption.  In experiments of CD$_4$:$^{18}$O$_2$ the actual rate constants for CD$_3^{18}$OD production are slightly lower than for $^{13}$CH$_3$OH; this could be due to deuterium exchange altering the apparent kinetics, or a difference in zero-point energy impacting the reaction rates.  As for $^{13}$CH$_4$:$^{16}$O$_2$ experiments there is no temperature dependence between 9K and 19K, indicating similar underlying energetics for the different isotopologue systems.  The lack of temperature dependence on the reaction rate is consistent with expectations from gas-phase and theoretical studies, which demonstrate barrierless or almost barrierless insertion of O($^1$D) into CH$_4$.

\newpage
\subsubsection{Other products}
\label{sec_resq}
\begin{figure*}
	\centering
	\includegraphics[width=0.9\linewidth]{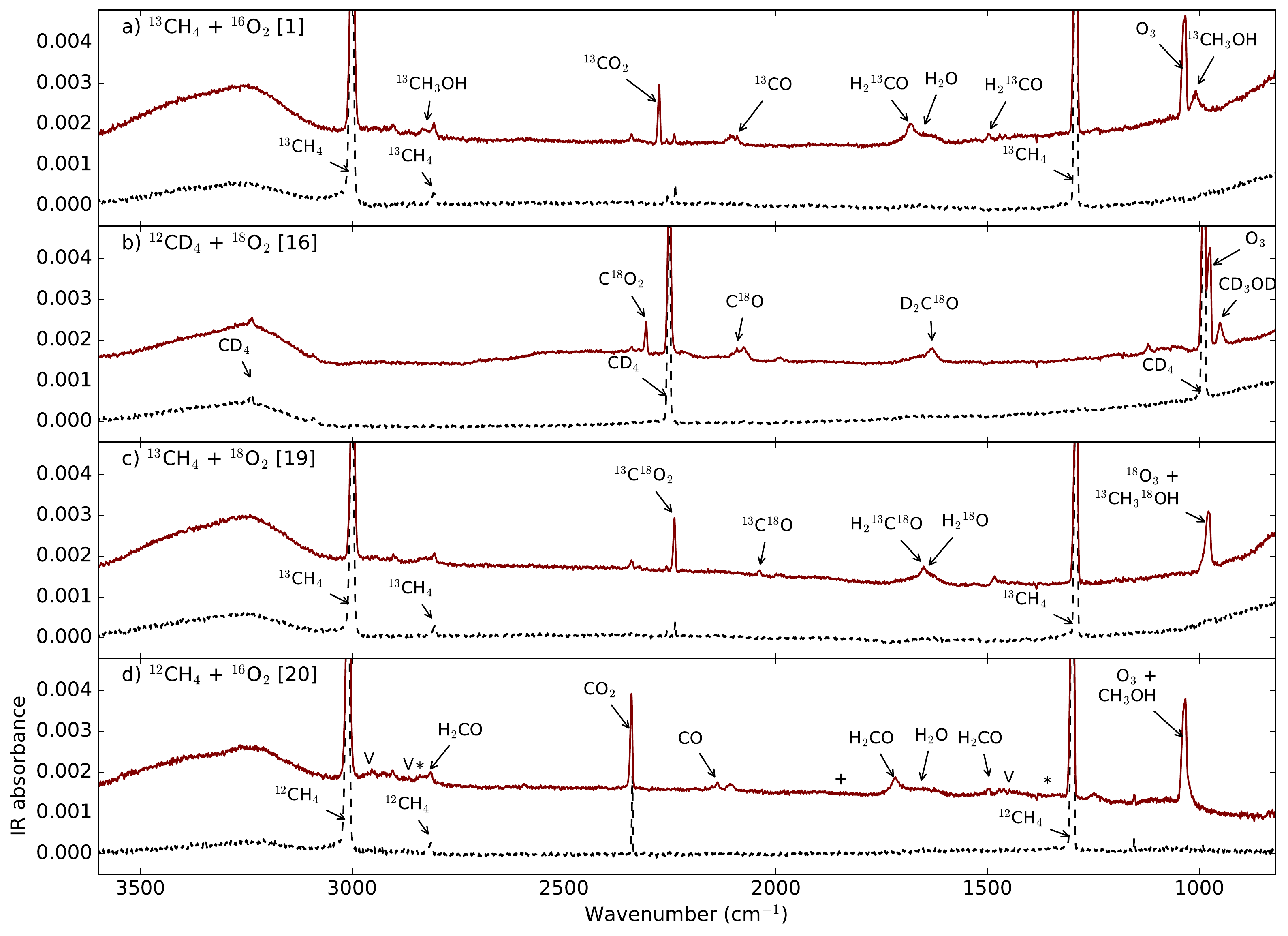}
	\caption{CH$_4$ + O$_2$ ice mixtures before (black dashed line) and after (red solid line) 2h UV irradiation, with reactant and product IR peaks labeled.  Each panel shows a different isotope combination, with the experiment number indicated in brackets.  In the $^{12}$CH$_4$:$^{16}$O$_2$ experiment (d), line centers for H$_2$O$_2$, C$_2$H$_6$, and HCO are shown with *'s, V's, and +'s respectively.}
	\label{ir_pr}
\end{figure*}

O$_3$, H$_2$CO, CO, CO$_2$, and H$_2$O are formed during the irradiation in addition to CH$_3$OH.  These are identified from their infrared features (Figure \ref{ir_pr}), which shift as predicted for different isotope combinations.  IR line centers used for identification and corresponding references are listed in Table 4. 

\begin{deluxetable*}{llcllc} 
	\tabletypesize{\footnotesize}
	\tablecaption{Observed infrared line centers for isotopologue experiments}
	\tablecolumns{6} 
	\tablewidth{\linewidth} 
	\tablehead{\colhead{Position (cm$^{-1}$)}             &
		\colhead{Assignment}							  &
		\colhead{Reference}                               & 
		\colhead{Position (cm$^{-1}$)}                    &
		\colhead{Assignment}							  &
		\colhead{Reference}                               }    
	\startdata
	\multicolumn{3}{l}{\underline{\textit{$^{13}$CH$_4$ + $^{16}$O$_2$}}}         & \multicolumn{3}{l}{\underline{\textit{$^{13}$CH$_4$ + $^{18}$O$_2$}}} \\           
	3000 & $^{13}$CH$_4$ 		  & a 										      &      3000 & $^{13}$CH$_4$  		  	& a   				\\
	2825 & $^{13}$CH$_{3}$OH      & a               						      &      2807 & $^{13}$CH$_4$  		  	& a   				\\
	2807 & $^{13}$CH$_4$ 		  & a 										      &      2240 & $^{13}$C$^{18}$O$_2$   	& e                    \\
	2276 & $^{13}$CO$_{2}$        & b               						      &      2039 & $^{13}$C$^{18}$O       	& f                    \\
	2092 & $^{13}$CO              & b               						      &      1652 & H$_2^{13}$C$^{18}$O    	& g                    \\
	1683 & H$_{2}^{13}$CO         & c*               						      &      1645 & H$_2^{18}$O            	& h*                   \\
	1652 & H$_2$O                 & b               						      &      1291 &  $^{13}$CH$_4$         	& a                    \\
	1498 & H$_{2}^{13}$CO         & c*               						      &       979 & $^{18}$O$_3$		   	& d*			        \\
	1291 & $^{13}$CH$_4$          & a               						      &      	 & $^{13}$CH$_3^{18}$OH   	& g                       \\
	1034 & O$_{3}$                & d*              						      &    \multicolumn{3}{l}{\underline{\textit{$^{12}$CH$_4$ + $^{16}$O$_2$}}}      \\  
	1008 & $^{13}$CH$_{3}$OH      & a               						      &      3009 &  CH$_4$                 & a                  \\
	\multicolumn{3}{l}{\underline{\textit{$^{12}$CD$_4$ + $^{18}$O$_2$}}}         &      2822 &  H$_2$CO                & b                  \\
	3237 & CD$_4$                & a                 						      &      2815 &  CH$_4$                 & a                  \\  
	2306 & C$^{18}$O$_2$         & e                 						      &      2341 &  CO$_2$                 & b                  \\
	2251 & CD$_4$                & a                 						      &      2139 &  CO                     & b                  \\
	2088 & C$^{18}$O             & f                 						      &      1718 &  H$_2$CO                & b                  \\
	1631 & D$_2$C$^{18}$O        & g                 						      &      1659 &  H$_2$O                 & b                  \\
	989 & CD$_4$                 & a                 						      &      1495 &  H$_2$CO                & b                  \\
	975 & O$_3$                  & d*                						      &      1300 &  CH$_4$                 & a                   \\
	952 & CD$_3^{18}$OD          & g                 						      &      1033 &  O$_3$ 			        & d*                 \\ 
	& 			                 &                   						      &      	  &  CH$_3$OH		        & b                  \\
	\enddata
	\tablenotetext{*}{Original assignment in RAIRS}
	\tablenotemark{a}{Reference spectrum in this work; }
	\tablenotemark{b}{\citet{Bouilloud2015}; }
	\tablenotemark{c}{\citet{Kaiser2015}; }
	\tablenotemark{d}{\citet{Schriver-Mazzuoli1995}; }
	\tablenotemark{e}{\citet{Du2011}; }
	\tablenotemark{f}{\citet{Legay-Sommaire1982}; }
	\tablenotemark{g}{Assigned based on shifts from other isotopologues; }
	\tablenotemark{h}{\citet{Zheng2011}}
	\label{line_ids}
\end{deluxetable*}

\begin{figure}
	\centering
	\includegraphics[width=\linewidth]{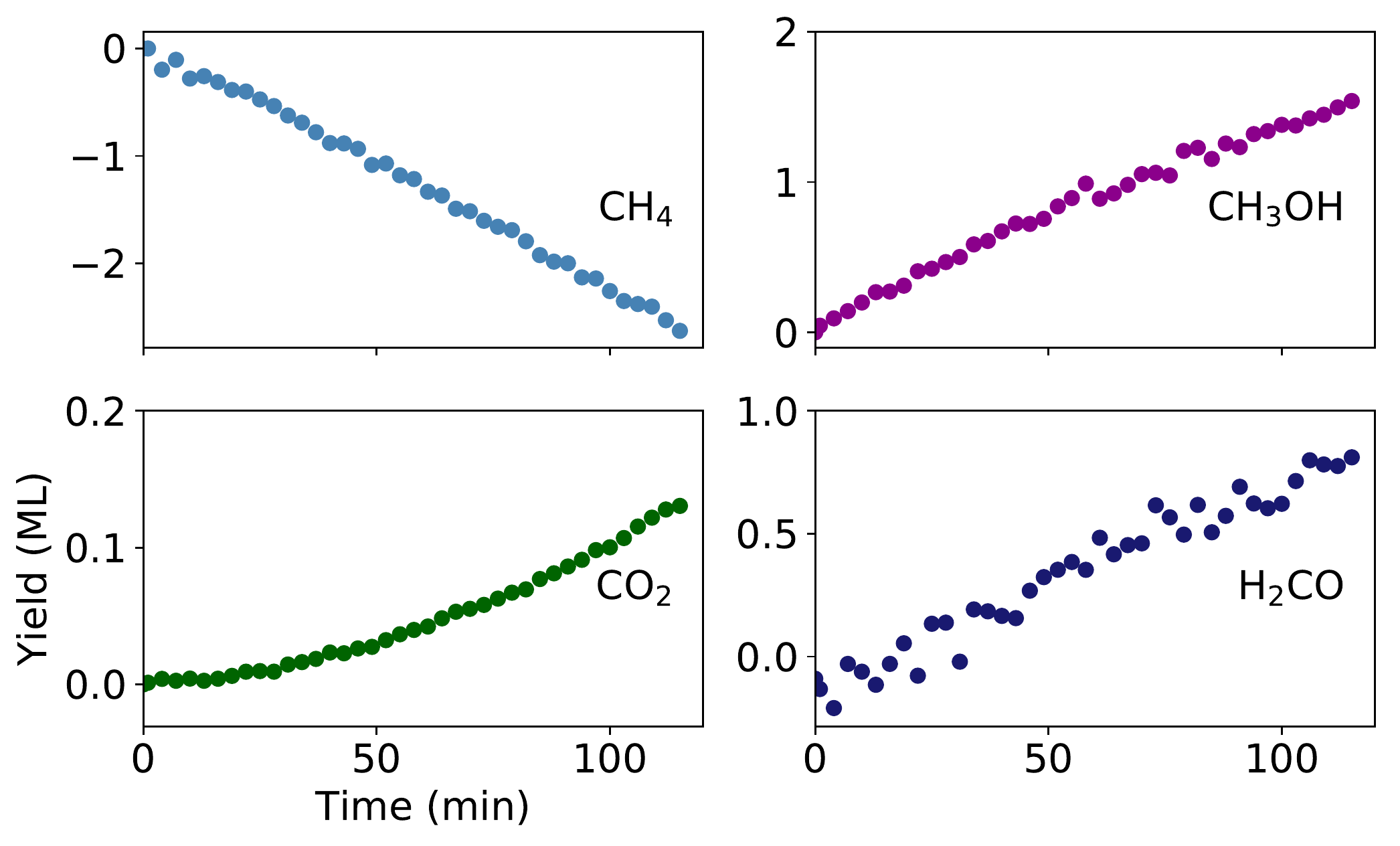}
	\caption{Growth curves in a $^{13}$CH$_4$:$^{16}$O$_2$ experiment (Exp. 3) for molecules with unblended IR features: CH$_4$ ($\nu_4$ bend), CH$_3$OH ($\nu_8$ CO str), CO$_2$ ($\nu_3$ a-str), and H$_2$CO ($\nu_3$ CH$_2$ sc).}
	\label{growthcurves}
\end{figure}

None of the isotopologue combinations result in clean IR features for all products.  For unblended features in the $^{13}$CH$_4$:$^{16}$O$_2$ experiments, growth curves can be measured from integrated IR spectra using the band strengths listed in Table 2.  An example set of growth curves is shown in Figure \ref{growthcurves}.  CH$_4$ is steadily consumed over the course of the experiment; CH$_3$OH and H$_2$CO growth begins at early times, while CO$_2$ growth accelerates later in the irradiation.

Final yields of all products are calculated from the IR spectrum after 2 hours of irradiation. For products with blended features, yields are estimated as follows.  The CO band is blended in $^{13}$CH$_4$:$^{16}$O$_2$ experiments, but not for $^{13}$CH$_4$:$^{18}$O$_2$ and $^{12}$CH$_4$:$^{16}$O$_2$ (Exps. 19 and 20).  In both unblended cases the ratio of CO/CO$_2$ is equal to $\sim$1.3.  Therefore, the CO yield in $^{13}$CH$_4$:$^{16}$O$_2$ experiments is estimated by multiplying the CO$_2$ yield by 1.3.  H$_2$O overlaps with the H$_2$CO feature at 1683cm$^{-1}$.  To estimate the H$_2$O yield, the H$_2$CO contribution is calculated based on the area of its 1498cm$^{-1}$ feature and then subtracted from the integrated area of the entire blended feature.

\begin{deluxetable*}{llllll} 
	\tabletypesize{\footnotesize}
	\tablecaption{Average product yields after 2h irradiation}
	\tablecolumns{6} 
	\tablewidth{\linewidth} 
	\tablehead{\colhead{CO}						        	 	&
		\colhead{CO$_2$}						         	&
		\colhead{H$_2$CO}						         	&
		\colhead{CH$_3$OH}						         	&
		\colhead{H$_2$O}						         	&
		\colhead{O$_3$}					        			\\
		\multicolumn{4}{c}{(\% wrt CH$_4$ consumed) }		&
		\multicolumn{2}{c}{(\% wrt O$_2$ dose)}             }
	\startdata
6.8 [0.6] & 5.3 [0.5] & 26.3 [8.7] & 60.0 [6.2] & 11.8 [4.5] & 12.7 [1.4] \\ 
	\enddata
	\tablenotetext{}{Standard deviations listed in brackets}
	\label{yields}
\end{deluxetable*}

Average product yields (Table 5) are calculated from experiments in Groups I, IV, and V, which all have sufficiently large IR features for all molecules to be quantified.  Total yields of carbon-bearing products are consistent with the measured consumption of CH$_4$ when considering band strength uncertainties. 

\subsubsection{Branching ratio}
\label{branch}
The ratios of carbon-bearing products can be used to derive the branching ratio of O($^1$D) + CH$_4 \rightarrow$ CH$_3$OH.  As seen in Table 5, CH$_3$OH has an average yield of 60$\pm$6\% with respect to CH$_4$ consumption.  This represents a lower limit to the CH$_3$OH formation efficiency since, as described in Section \ref{kinetics}, CH$_3$OH is also susceptible to photodissociation.  We can calculate the maximum abundance of carbon-bearing derivatives that may be formed from CH$_3$OH photoprocessing using the integrated rate law for product formation by Reaction \ref{rxn_mdiss}: 
\begin{equation}
\mathrm{[pr]}(t) = \frac{\mathrm{N_{ss}}[k_{dest}(1-e^{-k_rt}) - k_r(1-e^{-k_{dest}t})]}{k_{dest} - k_r},
\label{ratepr}
\end{equation}
where [pr] represents the combined abundance of products from CH$_3$OH destruction.  This represents an upper limit on the actual amount of products formed since not every photon absorption is necessarily dissociative.  Using the values of $k_r$ and N$_{\mathrm{ss}}$ derived from fitting CH$_3$OH growth curves (Table 3), we find an average upper limit of 9.7$\pm$1\% photo-products with respect to CH$_4$ consumption after 2 hours.

Assuming that CO and CO$_2$ are daughter species of H$_2$CO processing, the summed abundance of CO, CO$_2$, and H$_2$CO represents the total number of H$_2$CO molecules that are formed over the course of an experiment.  Experimentally, the average total abundance of H$_2$CO + CO + CO$_2$ is equal to 38$\pm$9\% with respect to CH$_4$ consumption.  Therefore, we observe more carbon-bearing side products than can be produced purely through CH$_3$OH photoprocessing.  The remaining products are likely formed from the product channel O($^1$D) + CH$_4$ $\rightarrow$ H$_2$CO + H$_2$.  The branching ratio of this channel should be equal to the difference between the observed side products and calculated abundances of photo-products.  This value will represent a lower limit since the calculated photoproduct yield is based on the assumption that every photon absorbed by CH$_3$OH is dissociative.  We obtain a lower limit for the H$_2$CO channel branching ratio of 29$\pm$9\%.  Based on this analysis we can therefore bracket the branching ratio of the CH$_3$OH channel between $\sim$60-71\%, with the lower limit representing a scenario with no CH$_3$OH photodissociation and the upper limit representing the maximum possible CH$_3$OH photodissociation.

Performing this same treatment at other time points, we find that the mean H$_2$CO branching ratio begins small and increases over the irradiation, flattening out to 28-29\% after 90 minutes.  This is consistent with a scenario in which some H$_2$CO comes directly from O insertion, and some from CH$_3$OH dissociation.

We note that an abstraction pathway of O + CH$_4$ $\rightarrow$ OH + CH$_3$ is a negligible or non-existent channel: if important, there should be considerable amounts of C$_2$H$_6$ and H$_2$O$_2$ from CH$_3$ + CH$_3$ and OH + OH, and we do not detect C$_2$H$_6$ or H$_2$O$_2$ as products.  We determine upper limits for each species using Experiment 20 ($^{12}$CH$_4$:$^{16}$O$_2$), as band strengths and positions are most reliable for the standard isotopes.  The line centers where H$_2$O$_2$ and C$_2$H$_6$ should appear are shown in Figure \ref{ir_pr}d; qualitatively, it is clear that these are minor species if they are present at all. To determine the C$_2$H$_6$ upper limit, we use the $\nu_{10}$ band at 2972cm$^{-1}$ \citep[2.2 x 10$^{-17}$cm molec$^{-1}$; ][]{Hudson2014}.  We find an upper limit of 0.05ML, or $\sim$1.7\% with respect to CH$_4$ consumption.  Only RAIRS band strengths are available in the literature for H$_2$O$_2$ and so, as for described for O$_3$ in Section \ref{quant}, we scale the RAIRS band strength reported in \citet{Loeffler2006}.  For the $\nu_6$ bending mode at 1368cm$^{-1}$ this yields a band strength of 3.0 x 10$^{-17}$cm molec$^{-1}$.  We find an H$_2$O$_2$ upper limit of 0.01ML, compared to a typical O$_3$ yield of $\sim$4ML for experiments with similar initial doses.  Thus, these species are either not produced or are a small fraction of the total reaction.

\subsection{CO dilution experiments}
\label{CO}
In order to evaluate whether CH$_3$OH formation via O($^1$D) insertion into CH$_4$ is also efficient in a more astrophysically realistic environment, we performed a set of experiments (21-26) in which the reactants $^{16}$O$_2$ and $^{13}$CH$_4$ were diluted in $^{12}$CO.  CO absorbs UV radiation, thereby introducing the possibility of contamination from CO-induced chemistry.   However, by following the formation of $^{12}$C vs. $^{13}$C products we determine that $^{12}$CO mainly reacts to form $^{12}$CO$_2$.  We follow the growth of $^{13}$CH$_3$OH in order to determine the oxygen insertion kinetics.

\begin{deluxetable}{lll}
	\tabletypesize{\footnotesize}
	\tablecaption{CH$_3$OH growth curve parameter fits: CO:O$_2$:CH$_4$ experiments}
	\tablecolumns{3} 
	\tablewidth{0.75\linewidth} 
	\tablehead{\colhead{Expt.}                           	&
		\colhead{$\mathrm{N_{ss}}$ (ML)}							         	&
		\colhead{$k_r$ (s$^{-1}$)}					        }
	\startdata
21 & 0.54 [0.15] & 1.1 [0.3] x 10$^{-4}$\\ 
22 & 0.70 [0.33] & 5.1 [2.6] x 10$^{-5}$\\ 
23 & 0.38 [0.15] & 9.8 [3.9] x 10$^{-5}$\\ 
24 & 1.43 [0.90] & 1.6 [1.3] x 10$^{-5}$\\ 
25 & 0.54 [0.45] & 3.7 [3.1] x 10$^{-5}$\\ 
26 & 0.55 [0.46] & 3.5 [3.1] x 10$^{-5}$\\ 
\enddata
	\tablenotetext{}{Uncertainties listed in brackets}
	\label{rates}
\end{deluxetable}

\begin{figure}
	\centering
	\includegraphics[width=0.8\linewidth]{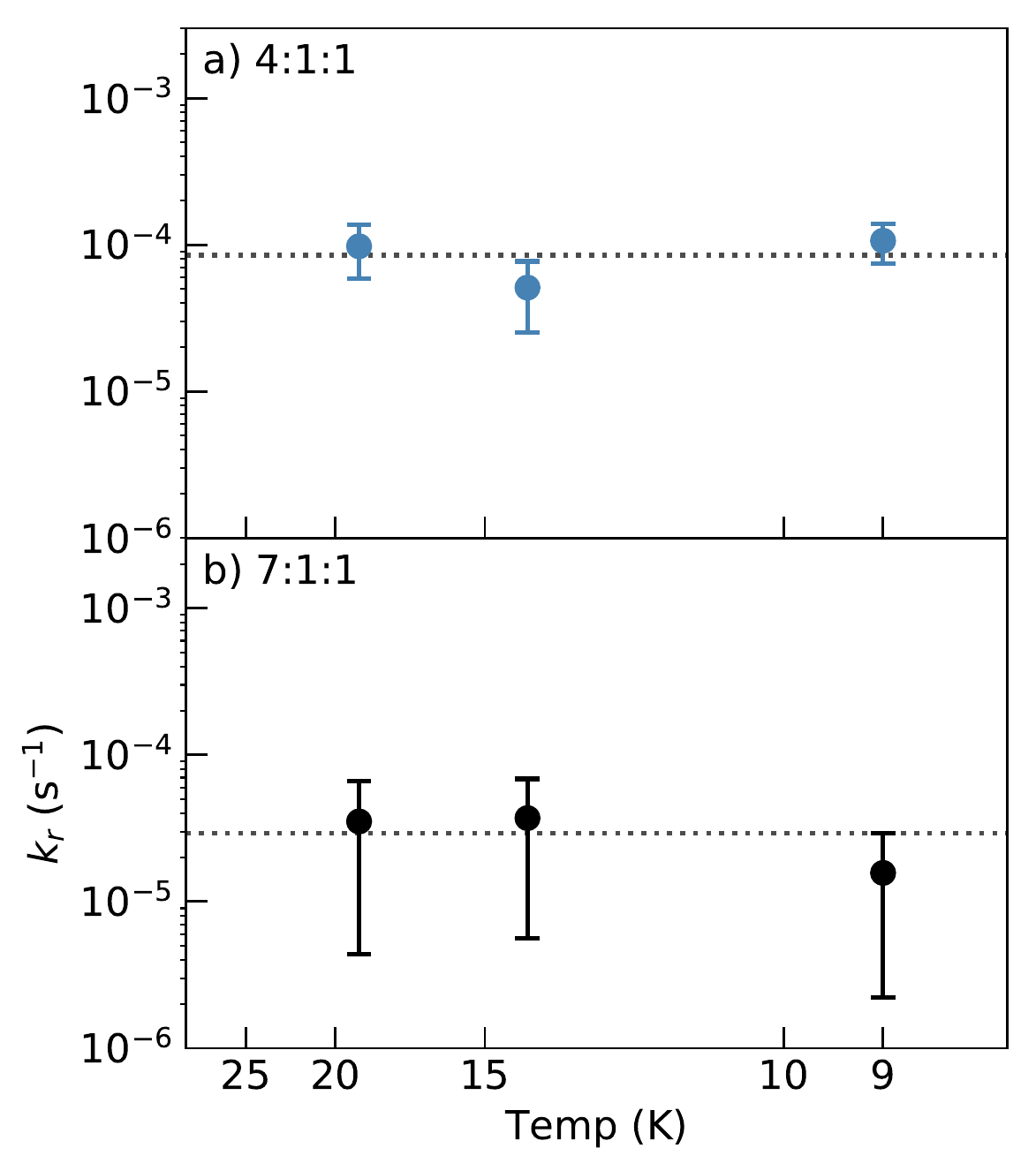}
	\caption{Rate constants as a function of temperature for CO dilution experiments.  a) 4:1:1 $^{12}$CO:$^{16}$O$_2$:$^{13}$CH$_4$.  b) 7:1:1 $^{12}$CO:$^{16}$O$_2$:$^{13}$CH$_4$.}
	\label{CO_mix}
\end{figure}

Rate constants for CH$_3$OH formation are determined by fitting Equation \ref{rateloss} as before.  CH$_3$OH formation rate constants are derived at temperatures between 9K-19K for ices of two different compositions: 4:1:1 $^{12}$CO:$^{16}$O$_2$:$^{13}$CH$_4$ mixtures and 7:1:1 $^{12}$CO:$^{16}$O$_2$:$^{13}$CH$_4$ mixtures.   Since product yields are reduced when the reactants are diluted in CO, uncertainties on the rate constants are higher than in the O$_2$:CH$_4$ experiments.  Still, as can be seen from the Arrhenius plots (Figure \ref{CO_mix}), there is no temperature dependence to the CH$_3$OH formation rate.  The parameters $k_r$ and N$_{ss}$ are shown in Table 6.

\section{Discussion}
\label{sec_discussion}
\subsection{Reaction network}
\label{netwk}
\begin{figure*}
	\centering
	\includegraphics[width=0.9\linewidth]{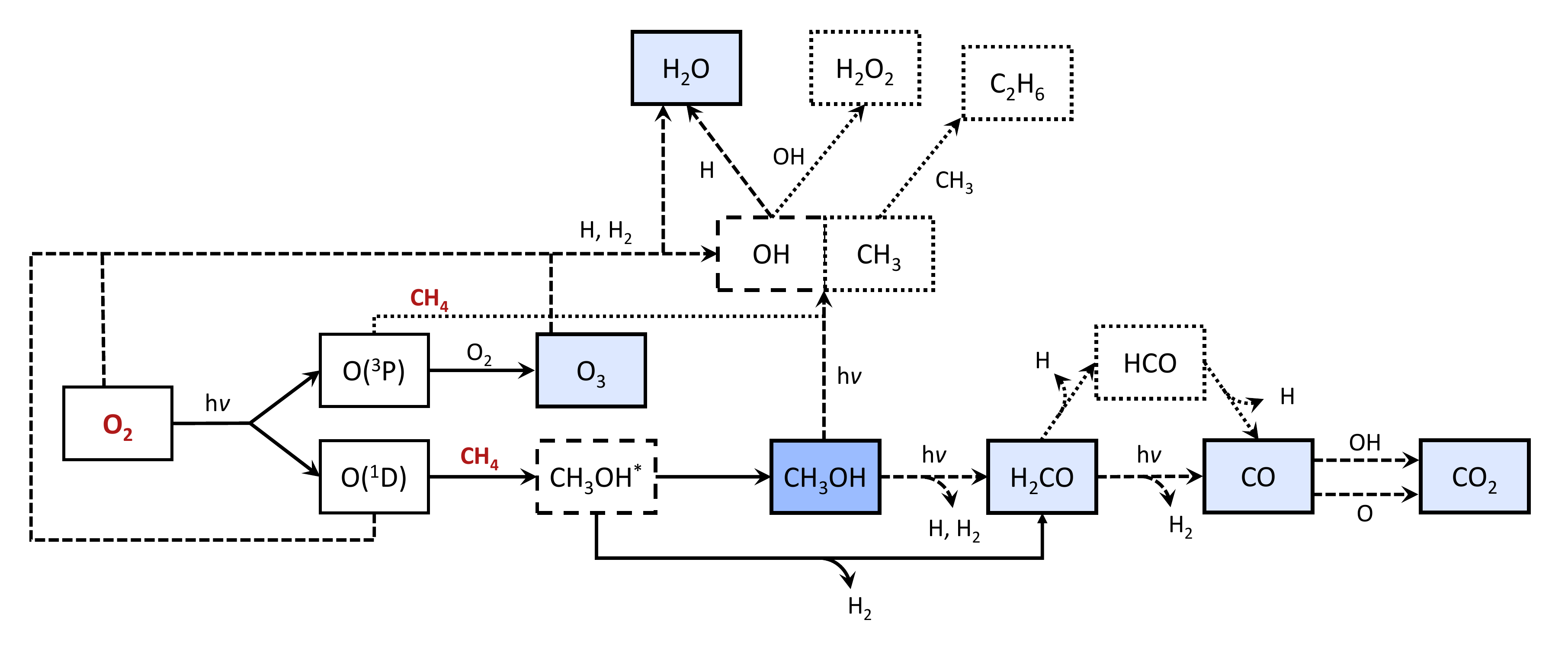}
	\caption{Proposed network for this reaction system.  Reactants are shown with red text, and observed products as blue boxes.  Boxes with dashed borders indicate intermediates that are likely based on observed products; dotted borders indicate intermediates/products that are not observed and therefore minor channels if present at all.  Solid lines represent pathways leading to primary products; dashed lines represent pathways leading to secondary products; dotted lines represent pathways that are likely minor contributions based on the lack of observed products.}
	\label{network}
\end{figure*}

As mentioned in Section \ref{sec_intro}, for the UV wavelengths in this study, O$_2$ dissociation proceeds through the channel O$_2$ $\rightarrow$ O($^1$D) + O($^3$P) with an efficiency of unity \citep{Lee1977}.  Thus, our ices should contain equal parts O($^1$D) and O($^3$P).  We now draw from gas-phase, theoretical, and, when available, condensed-phase studies to interpret the formation pathways of our observed products.  Figure \ref{network} presents a summary of the possible pathways we discuss.

\textit{CH$_3$OH formation}:  In the gas phase the insertion of O($^1$D) into CH$_4$ results in an excited CH$_3$OH product, which undergoes unimolecular dissociation to form mainly OH + CH$_3$ unless it is stabilized by e.g. collision \citep{DeMore1967,Parnis1993} or supersonic expansion \citep{Hays2015}.  As our experiments involve condensed ices, it is fully consistent that the intact CH$_3$OH molecule is observed due to energy dissipation into the solid.  This insertion process has been demonstrated to have essentially no barrier in gas-phase and theoretical studies \citep{DeMore1967,Yu2004}, which is again consistent with the lack of a temperature dependence to CH$_3$OH formation in our experiments.  

In contrast, ground state O($^3$P) oxygen atoms follow an abstraction channel with CH$_4$ to produce OH + CH$_3$, with an estimated barrier of over 5000K \citep[e.g.][]{Walch1980,Zhao2016}.  Such an abstraction channel could conceivably lead to CH$_3$OH production through radical recombination, however this is unlikely due to the high theoretical energy barrier.  Furthermore, as stated in Section \ref{branch}, the upper limits for C$_2$H$_6$ and H$_2$O$_2$ are on the order of a percent or less of the total consumed CH$_4$, indicating that abstraction is not an important process in this system.

\textit{O$_3$ formation}:  O$_3$ formation from energetic processing of molecular oxygen under astrochemically relevant conditions has been well-described in the literature \citep[e.g.][]{Schriver-Mazzuoli1995,Bennett2005a,Sivaraman2011}.  The mechanism for O$_3$ formation under these conditions is O($^3$P) + O$_2$.  This is likely the formation pathway occurring in our ices as well since O($^{3}$P) should be formed in similar quantities as O($^1$D), and the barrier to react with CH$_4$, the other available reaction partner, is high.

\textit{H$_2$CO formation}:  A fraction of H$_2$CO may be formed from photo-processing of CH$_3$OH, as has been demonstrated in previous UV irradiation studies \citep[e.g.][]{Gerakines1996,Oberg2009a}.  As discussed in Section \ref{branch}, the majority of H$_2$CO is likely formed directly from insertion of O($^1$D) into CH$_4$.  Indeed, following CH$_3$OH, H$_2$CO is the next most stable possible product of O($^1$D) insertion into CH$_4$ \citep{Chang2002,Yu2004}, and gas-phase and matrix studies typically show H$_2$CO as the second-most common insertion product \citep[e.g.][]{DeMore1967,Appelman1989,Hays2015}.  Moreover, from Figure \ref{growthcurves} the shape of the H$_2$CO growth curve is different from the second-generation product CO$_2$, but quite similar to that of the primary insertion product CH$_3$OH.  This is consistent with a scenario in which the majority of H$_2$CO forms from the same mechanism as CH$_3$OH, as opposed to growing only from CH$_3$OH processing. 

\textit{CO formation}:  Successive H abstractions could form CO via H$_2$CO $\rightarrow$ HCO $\rightarrow$ CO, but as seen in Figure \ref{ir_pr}d HCO is not observed in the IR despite having a comparable band strength to CO \citep{Bennett2007a}.  Alternatively, CO may be produced by unimolecular dissociation of H$_2$CO to directly form CO + H$_2$, which is a demonstrated photo-process of H$_2$CO in the condensed phase \citep{ThomasJr.1973} and explains the lack of observed HCO.

\textit{H$_2$O formation}:  Hydrogenation channels beginning with O, O$_2$, or O$_3$ have been shown experimentally to lead to water formation \citep{Dulieu2010,Ioppolo2008,Romanzin2011} and are likely at play in this system given the availability of these reactants.  An alternative pathway is the formation and subsequent hydrogenation of OH by the abstraction pathway of O($^3$P) with CH$_4$.  However, given the high barrier to the O($^3$P) abstraction pathway and lack of observed C$_2$H$_6$, reaction pathways beginning with O, O$_2$, or O$_3$ are more likely.

\textit{CO$_2$ formation}:  From the shape of its growth curve (Figure \ref{growthcurves}), CO$_2$ is almost certainly a second-generation product, with its abundance increasing at later times in the experiment.  One possible route for CO$_2$ formation is CO + OH, with OH formed as an intermediate along the H$_2$O formation channels.  Studies have also shown CO$_2$ formation from CO + O \citep{Parnis1993,Madzunkov2006,Minissale2013}.  It is unclear whether the higher energy barrier of the CO + O pathway or the higher diffusion barrier of the CO + OH pathway will dominate under these conditions.  

\subsection{O insertion reaction}
\label{mech}
We derive a branching ratio for the CH$_3$OH channel of $\sim$65\%, with the remaining insertions leading to H$_2$CO formation.  Fragmentation does not appear to be an important process in this system as evidenced by the lack of CH$_3$ and OH chemistry.  Solid-state oxygen insertions should therefore lead to a net increase in chemical complexity, in most cases increasing the size of the product molecule and in all cases forming an O-containing organic from a hydrocarbon.

We observe no temperature dependence to CH$_3$OH formation via oxygen insertion, consistent with gas-phase and theoretical studies which show a negligible or non-existent energy barrier.  With these experiments alone, however, we cannot definitively rule out the presence of a small energy barrier for several reasons.  First, our experiments cannot isolate any contributions from tunneling, which may play a role at the low temperatures studied here.  Additionally, it is possible that a diffusion barrier for O atoms could mask an insertion barrier, since the atoms would have to overcome a barrier to diffuse away than to react.  Based on rather low diffusion barriers for O atoms in models \citep[$\sim$400K; ][]{Garrod2011}, only a small reaction barrier could be masked in this way.  Finally, the photodissociation of O$_2$ may lead to the formation of ``hot'' oxygen atoms: the threshold for formation of O($^3$P) + O($^1$D) is 175nm \citep{Nee1997}, whereas the UV lamp in this study peaks at 160.8nm.  This energy difference represents $\sim$0.6eV.  Some of this excess energy will be dissipated into the solid, but it is possible that oxygen atoms formed as a result of photodissociation are superthermal.

A hot atom mechanism may also be at play in ISM ices: as described in Section \ref{sec_intro}, the O($^1$D) required for oxygen insertion chemistry in astrophysical settings is likely formed from photolysis or radiolysis of oxygen-bearing molecules.  The possibility that oxygen insertion is driven by hot atoms in our experiments makes it important to explore in ices with more realistic compositions than O$_2$:CH$_4$ mixtures.  In the CO-dominated experiments (Section \ref{CO}), we found that even when the reactants are diluted in a CO ice, the CH$_3$OH formation rate follows the same temperature-independent trend as in the O$_2$:CH$_4$ only experiments.  Thus, whether O($^1$D) insertion is mediated by a hot atom mechanism or not, we find that it can proceed at very low temperatures in a barrierless/pseudo-barrierless manner.

When taken along with the theoretical insertion barrier of just $\sim$280K and a negligible measured insertion barrier in the gas phase, an essentially barrierless ice-phase insertion process is the most likely explanation for our experimental results.  We conclude that insertion is a dominant reaction pathway when excited O($^1$D) is present, and furthermore that it proceeds pseudo-barrierlessly in ISM-like ices.

\subsection{Astrophysical implications}
In the interstellar medium, O insertion pathways could be of particular importance in very cold regions where radical diffusion chemistry is not thermally accessible.  Indeed, gas-phase chemistry leading to hydrocarbon formation is known to be very efficient at low temperatures; these hydrocarbons can then accrete onto grain surfaces.  Meanwhile, as mentioned in Section \ref{sec_intro}, excited O($^1$D) atoms can be formed by energetic processing of common oxygen-bearing constituents of astrophysical ices (in particular, H$_2$O and CO$_2$).  The insertion of oxygen atoms directly into hydrocarbons would then lead to the formation of a variety of complex organic species, without the need for radical diffusion.  The degree to which this type of chemistry contributes to COM formation will need to be tested by astrochemical modelers.  In this case, it will be important to distinguish between O($^3$P) and O($^1$D) atoms in order to accurately account for the chemistry.

Ultimately it will be important to quantify the process of O insertion into larger hydrocarbons:  CH$_3$OH, while a convenient test case, can be produced in astrochemical models via CO hydrogenation.  Larger organics, on the other hand, are regularly underproduced in models and therefore seemingly missing a formation pathway.  From the limited literature available, O insertions into both saturated and unsaturated larger hydrocarbons should occur at low temperatures in condensed phases \citep{Parnis1993,DeMore1969}, though further experiments are required to quantify the energetics and product distribution of such systems.  While we expect to see a low or non-existent insertion barrier for larger hydrocarbons, it is important to obtain branching ratios for the insertion products: the C-H bonds are not necessarily degenerate in larger hydrocarbons as they are in CH$_4$, making it difficult to predict a priori which products will form.  Moreover, gas-phase studies suggest that an abstraction channel becomes increasingly competitive with insertion for larger hydrocarbons \citep{Luntz1980}.  There is also some evidence for O insertion into C-C bonds in the gas phase \citep{Yang2006}.  The experimental setup used in this study cannot be used to test oxygen insertions into other hydrocarbons, as most larger hydrocarbons have appreciable UV absorption cross-sections above the sapphire window cutoff; however, testing these systems experimentally should be possible with an atomic beam setup.  Quantitative constraints on the energetics and product distributions of O atom insertion into larger hydrocarbons will enable an evaluation of the importance of oxygen insertion chemistry to forming complex molecules in astrophysical environments.

\section{Conclusions}
We have experimentally tested and quantified the formation of CH$_3$OH via oxygen insertion into methane in astrophysical ice analogs.  From our results we conclude:
\begin{enumerate}
	\item Selective dissociation of O$_2$ in mixed O$_2$:CH$_4$ ices results in the formation of CH$_3$OH in various isotopologue studies.  The growth kinetics of CH$_3$OH are well described by a model that includes both formation and photo-dissociative loss.
		\item A direct insertion mechanism of O($^1$D) atoms into CH$_4$ explains CH$_3$OH formation, with a minor channel of H$_2$CO production and no evidence of fragmentation to CH$_3$ + OH.  O($^1$D) insertions in ices therefore lead to a net increase in chemical complexity.  We quantify the steady-state branching ratio to CH$_3$OH to be 60-71\%.
	\item Experiments with varying ice thicknesses and reactant ratios show no temperature dependence to the CH$_3$OH formation rate constant $k_r$ from 9-24K.  This holds even when the reactants are diluted in a CO matrix, consistent with a small or non-existent energy barrier to insertion.
	\item Experimental constraints on the energetics and branching ratios of O insertions into larger hydrocarbons are required to assess the contribution of oxgyen insertion chemistry to observed abundances of COMs.
\end{enumerate}  

\noindent The authors thank Edith Fayolle, Robin Garrod, and Ilsa Cooke for valuable feedback.  J.B.B acknowledges funding from the National Science Foundation Graduate Research Fellowship under Grant DGE1144152. K.I.O. acknowledges funding from the Simons Collaboration on the Origins of Life (SCOL) investigator
award.

\software{emcee (Foreman-Mackey et al. 2013)}

\section{Appendix A: IR spectrum fitting with MCMC}
\begin{figure}[h!]
   \centering
	\includegraphics[width=\linewidth]{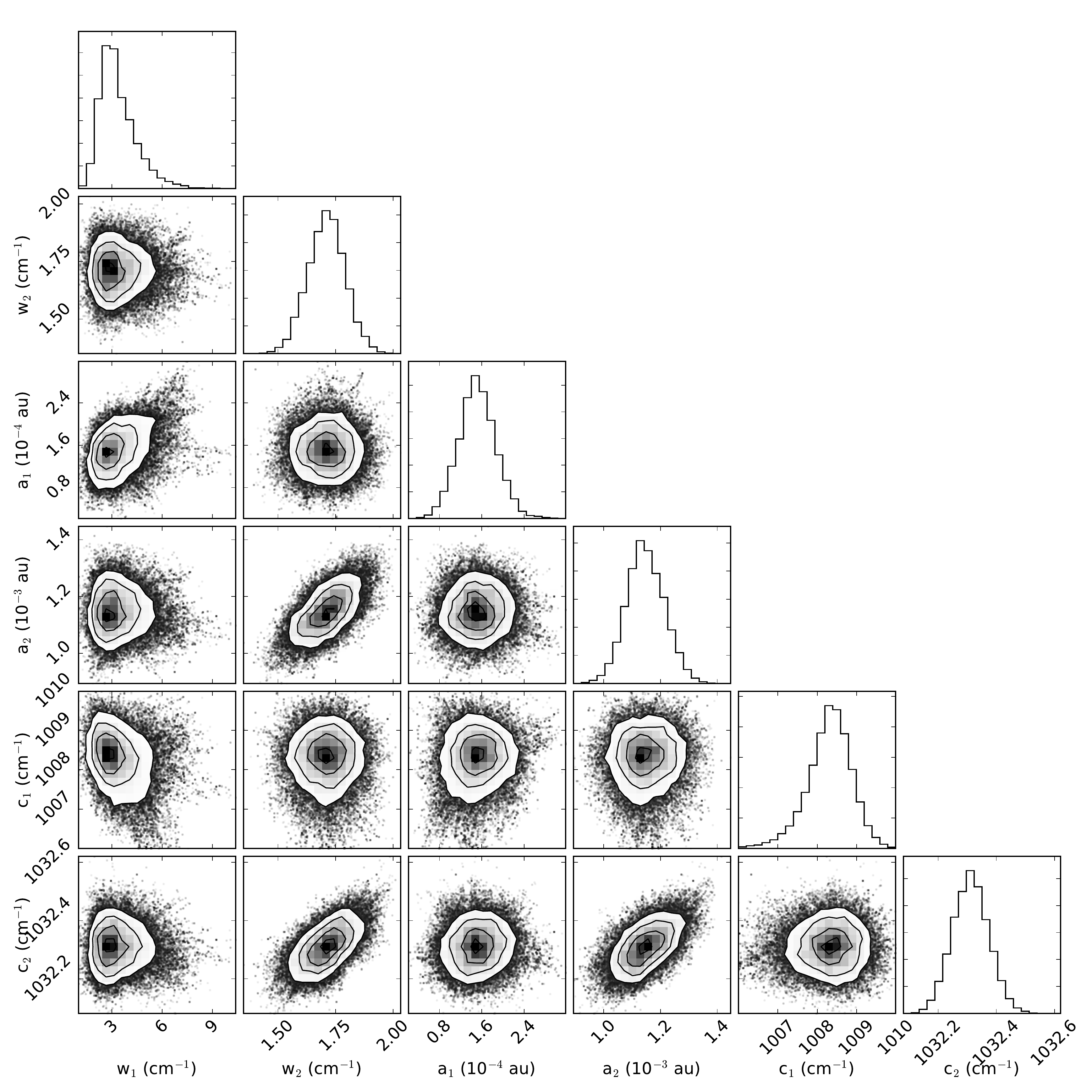}
	\caption{Example corner plot showing the covariance of two Gaussians used to fit the spectrum shown in Figure \ref{fits}a. }
	\label{corner}
\end{figure}

For $^{16}$O$_2$:$^{13}$CH$_4$ experiments, each IR spectrum was fit from 940-1090cm$^{-1}$.  The scan immediately prior to irradiation was subtracted from each irradiation spectrum.  Irradiation spectra were then fit with a model consisting of a linear baseline term and four Gaussians:
\begin{multline}
y(x) = y_{o} + m(x - x_{o}) + a_1e^{-(x-b_1)^2/2c_1^2} + \\ a_2e^{-(x-b_2)^2/2c_2^2} + a_3e^{-(x-b_3)^2/2c_3^2} + a_4e^{-(x-b_4)^2/2c_4^2},
\end{multline}
where $y_{o}$ and $x_{o}$ represent the y and x offsets for the baseline, $a_n$ are the Gaussian amplitudes, $b_n$ are the Gaussian centers, and $c_n$ are the Gaussian widths.  An example corner plot from MCMC fitting with $emcee$ is shown in Figure \ref{corner}; for clarity, the fit parameters for two Gaussians are shown rather than all 16 parameters simultaneously.  For fitting the $^{18}$O$_2$:CD$_4$ experiments, the wavelength range used for fitting was 900-965cm$^{-1}$, and only two Gaussians were required in the model since the $^{18}$O$_3$ feature is sufficiently well-separated from CD$_3^{18}$OD.  For CO:$^{16}$O$_2$:$^{13}$CH$_4$ experiments, spectra were fit from 980-1030cm$^{-1}$ and due to low product yields a single Gaussian was sufficient to fit the CH$_3$OH feature.

\section{Appendix B: Experimental growth curves and model fits}
\begin{figure}[h!]
   \centering
	\includegraphics[width=\linewidth]{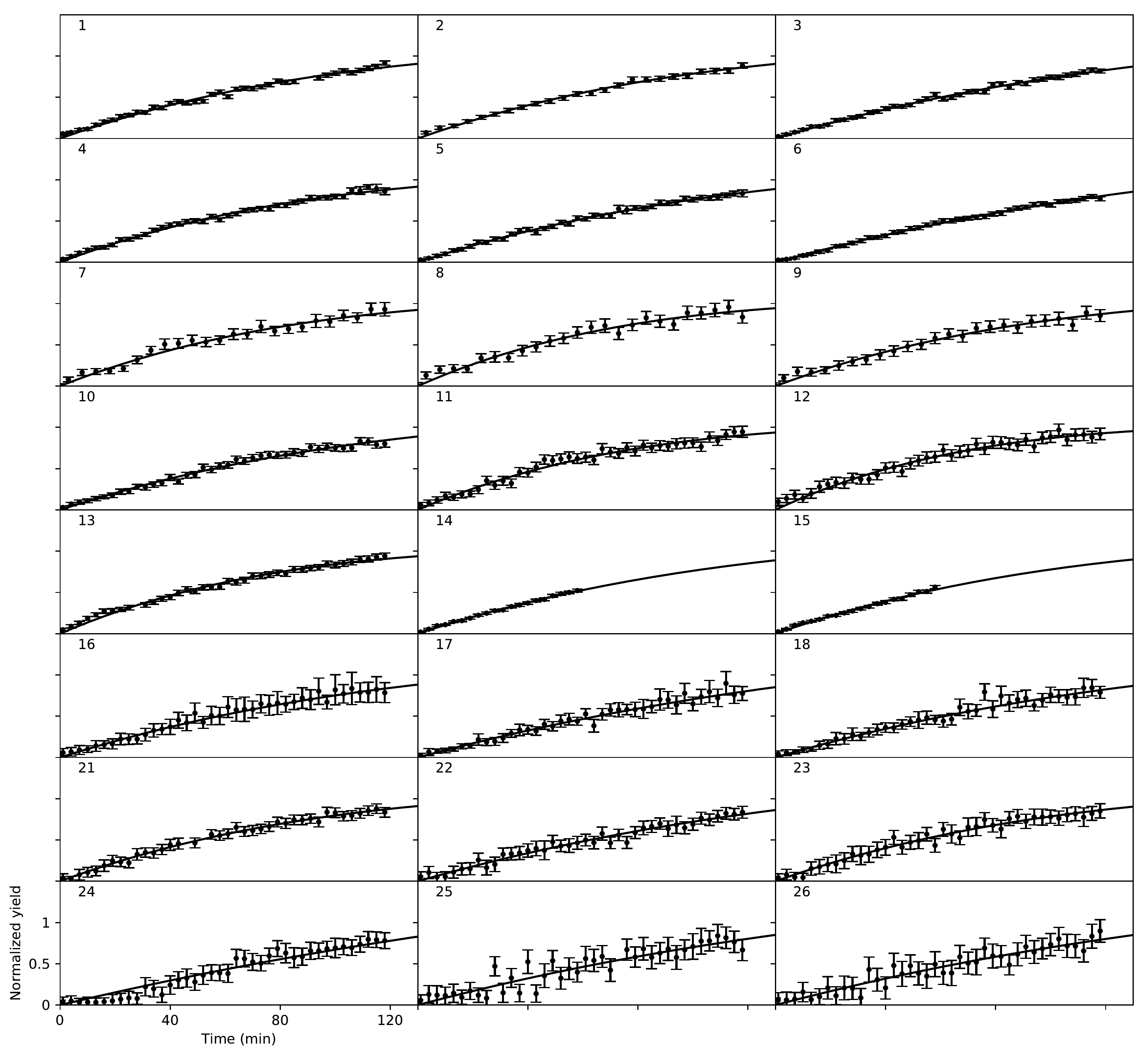}
	\caption{Experimental growth curves with best-fit kinetic models for all experiments.  Experiment numbers from Table 1 are listed in each subplot.}
	\label{gcurves}
\end{figure}

Growth curves along with best-fit kinetic models for all experiments are shown in Figure \ref{gcurves}.

\newpage
\bibliographystyle{apj}

\end{document}